\documentclass[twocolumn, times, twocolappendix]{aastex701}
\usepackage{amsmath}
\usepackage{hyperref}

\def\IHEP{State Key Laboratory of Particle Astrophysics, Institute of High Energy Physics, Chinese Academy of Sciences, 19B Yuquan Road, Beijing 100049, People’s Republic of China; \href{mailto:liyanrong@mail.ihep.ac.cn}{liyanrong@mail.ihep.ac.cn}}
\begin{document}
\title{On the Probability Distribution and Null-hypothesis Testing of Cross-correlation for Light Curves in Active Galactic Nuclei}
\author[0000-0001-5841-9179]{Yan-Rong Li}
\affiliation{\IHEP}
\email{liyanrong@mail.ihep.ac.cn}
\author[0000-0001-9449-9268]{Jian-Min Wang}
\affiliation{\IHEP}
\affiliation{School of Astronomy and Space Sciences, University of Chinese Academy of Sciences, Beijing 100049, People’s Republic of China}
\affiliation{National Astronomical Observatory of China, 20A Datun Road, Beijing 100020, People’s Republic of China}
\email{wangjm@mail.ihep.ac.cn}

\begin{abstract}
Cross-correlation is crucial in studies of multiwavelength flux variability in active galactic nuclei (AGNs),
especially for reverberation mapping
analysis, where interpolated cross-correlation function is widely used to measure time lags between light curves.
While time-lag uncertainties can be estimated via the flux
randomization and random subset selection method,
an appropriate framework for assessing cross-correlation significance remains lacking in the literature.
Here we attempt to fill this gap by leveraging the well-established property from stochastic time series theory, namely that
the probability distribution of cross-correlation coefficients for independent stochastic light curves asymptotically
approaches a normal distribution. Its variance can be analytically estimated using the auto-correlation functions
of the light curves. We employ Monte Carlo simulations to validate this property
for irregularly sampled, red-noise AGN light curves, and then propose a fast procedure to perform
null-hypothesis testing for the cross-correlation of AGN light curves. We also present exemplary applications to AGN
reverberation mapping data. This procedure requires prior determination of the auto-correlation functions
of light curves, which can be obtained via model fitting. The long-standing issue
regarding unbiasedly recovering the auto-correlation function remains unresolved
when light-curve duration is comparable to the typical variation timescale, warranting further future
investigation.
\end{abstract}

\section{Introduction}
Active galactic nuclei (AGNs) are known to exhibit flux variability across the entire electromagnetic spectrum
(\citealt{Mushotzky1993, Ulrich1997}). Such variability provides profound insights into the structure and
kinematics of the various components in AGNs (\citealt{Antonucci1993, Netzer2015}), most of which are spatially
compact on cosmic scales and cannot be directly resolved with current observational facilities.
One important application of AGN flux variability is the so-called reverberation mapping (RM).
Its concept was first proposed by \cite{Bahcall1972} and its mathematical framework was subsequently
established by \cite{Blandford1982}. Early RM observations primarily focused on broad emission lines
in AGNs, aimed at probing geometry and kinematics of broad-line regions (\citealt{Peterson1988, Peterson1993}), which
serve as a key component for measuring masses of supermassive black holes residing at the centers of AGNs (\citealt{Peterson2014}).
Shortly afterwards, RM was further extended to diagnosing accretion disks and dusty tori  in AGNs
(e.g., \citealt{Clavel1989, Nelson1996, Collier1998, Collier1999}).

A fundamental procedure involved in RM is to conduct cross-correlation analysis between light curves across different wavelength
bands to determine their inter-band time lags. A widely used analysis approach is calculating the interpolated cross-correlation
function (ICCF), in which the peak cross-correlation coefficient is determined by searching over
a time-lag range and the time lag between light curves is assigned by the peak location or the centroid of the ICCF above
a predefined threshold fraction of the ICCF peak value (\citealt{Gaskell1987, Peterson1998}; see also \citealt{Edelson1988}
and \citealt{Alexander1997} for discrete cross-correlation function).
The uncertainties in the measured time lag are estimated using the flux randomization and
random subset selection (FR/RSS) method (\citealt{Peterson1998}).
Although this approach has been well established, it does not address the question of whether the light curves
are correlated. In this respect, the literature still lacks a proper, unified framework
for assessing reliability of the cross-correlation and performing the corresponding null-hypothesis testing.
For independent white-noise light curves, the probability distribution of cross-correlation coefficients
is known analytically (\citealt{Fisher1921}; see also \citealt[Chapter XVI]{Uspensky1937}). Unfortunately,
AGN variability is typically characterized by red noise (\citealt{Kelly2009}). It is not trivial to extend the analytical
probability distribution derived under the white-noise assumption to the red-noise regime.

Previous studies commonly employ Monte Carlo simulations to evaluate the significance of
cross-correlation, without investigating the probability distributions of
cross-correlation coefficients (e.g., \citealt{Peterson1998,MaxMoerbeck2014, Li2021, U2022, Wang2024}).
Indeed, the time series theory has established that for any type of independent stochastic light curves,
the probability distribution of cross-correlation coefficients asymptotically approaches a normal distribution,
given that the light curves are described by stationary processes (\citealt[Section 11.2]{Brockwell1991}).
This can be regarded as a corollary of the central limit theorem.
The variance of the normal distribution can be estimated analytically using the auto-correlation functions
of the light curves, which are in turn obtained by fitting a prescribed variability model.
This work is devoted to applying the above property to AGN light curves and
developing a framework for null-hypothesis testing of cross-correlation
of AGN light curves in ICCF analysis.

This paper is organized as follows. Section~\ref{sec_ccf} demonstrates
that cross-correlation coefficients of uncorrelated, stochastic light curves follow normal distributions, where the variances
are quantitatively expressed with auto-correlation functions of light curves.
Section~\ref{sec_tests} performs a suite of simulation tests to verify the rationale of the normal distributions
for cross-correlation coefficients and validates the derived theoretical expressions for the variances.
Section~\ref{sec_null} describes a procedure for null-hypothesis testing of the cross-correlation for AGN light curves and
applies it to realistic RM data. The conclusion and discussion are summarized in Section~\ref{sec_conclusion}.

\section{Theoretical Analysis}\label{sec_ccf}
\subsection{The Cross-correlation Function}
\subsubsection{Mathematical Definitions}
The cross-correlation function between two light curves, say, $x(t)$ and $y(t)$, is defined as (e.g., \citealt{Welsh1999})
\begin{equation}\label{eqn_rtau}
r(\tau) = \frac{E\left\{[x(t)-\bar x][y(t+\tau)-\bar y]\right\}}{\sqrt{E\{[x(t)-\bar x]^2\}E\{[y(t+\tau)-\bar y]^2\}}},
\end{equation}
where $\tau$ is the time lag, $\bar x$ and $\bar y$ are the mean of $x(t)$ and $y(t+\tau)$,
and $E\{x\}$ represents the expected value of $x$.
As mentioned above, it has been established in time series theory that, for {\it stationary and uncorrelated}
light curves, the probability distribution of $r$, denoted as $p(r)$, asymptotically
approaches a normal distribution with a zero mean and a variance depending on the variation properties
of the light curves (\citealt[Section 11.2]{Brockwell1991}). Here, ``stationary'' means that the expectation and covariance
of the light curve are time-independent.

\cite{Fisher1921} found that the $z$ transformation, namely,
\begin{equation}
z = \frac{1}{2}\ln\left(\frac{1+r}{1-r}\right) = {\rm tanh^{-1}}(r),
\end{equation}
has a property that the probability distribution of $z$, denoted as $p(z)$, approaches the normal distribution
more rapidly than $p(r)$ even for a comparatively small number of data points. By noting that
\begin{equation}\label{eqn_rz}
z \approx r + \mathcal{O}(r^3),
\end{equation}
we expect that $p(z)$ approximately approaches the same normal distribution as $p(r)$.  Another advantage of using $z$ is that
the allowed range for $z$ is $(-\infty, \infty)$, which better accommodates
the normal distribution. In contrast, $r$ is restricted to $[-1, 1]$. Hereafter, unless stated otherwise, we
use $z$ to delineate cross-correlation coefficients.

\subsubsection{The General Case}
Let's directly express the probability $p(z)$ as the normal distribution with a zero mean and variance of $\sigma_z^2$
\begin{equation}
p(z) = \mathcal{N}(0, \sigma_z^2),
\end{equation}
where $\mathcal{N}$ represents the normal distribution.
The variance $\sigma_z^2$ is related to the auto-correlation coefficients of $x(t)$ and $y(t)$ as
(\citealt[Section 11.2.2]{Brockwell1991})
\begin{eqnarray}\label{eqn_norm}
\sigma_z^2 &=& \frac{1}{n} \left[1 + 2\sum_{k=1}^n\left(1-\frac{k}{n}\right)\rho_x(k) \rho_y(k)\right],
\end{eqnarray}
where $n$ is the number of points in the light curves, $\rho_x(k)$ and $\rho_y(k)$ are the auto-correlation
functions at a time lag of $k\Delta t$
between $x(t)$ and $y(t)$, respectively, and $\Delta t$ is the sampling interval of the light curves.
From Equation~(\ref{eqn_norm}), we can define an effective number of points
\begin{equation}\label{eqn_neff1}
n_{\rm eff} = n \left[1 + 2\sum_{k=1}^n\left(1-\frac{k}{n}\right)\rho_x(k) \rho_y(k)\right]^{-1},
\end{equation}
and the variance is then given by
\begin{equation}\label{eqn_neff2}
\sigma_z^2=\frac{1}{n_{\rm eff}}.
\end{equation}
In Appendix, we provide a simple derivation for Equation~(\ref{eqn_norm}).

\begin{figure}
\centering
\includegraphics[width=0.48\textwidth]{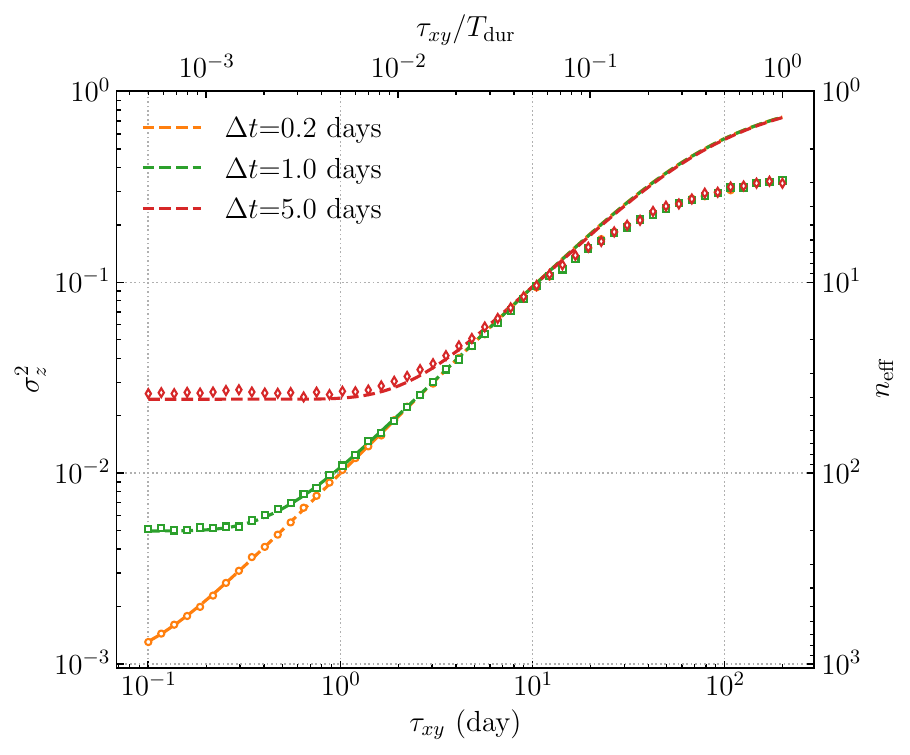}
\caption{The variance $\sigma_z^2$ and $n_{\rm eff}$ versus $\tau_{xy}$ for different sampling intervals $\Delta t$.
The dashed lines show the variance $\sigma_z^2$ calculated with Equations~(\ref{eqn_norm}) and (\ref{eqn_rho}).
The points show the variance $\sigma_z^2$ calculated from simulations, where the light curves are assumed to be
regularly sampled with a time duration of $T_{\rm dur}=200$ days (see the text in Section~\ref{sec_tests}).
The right vertical axis shows $n_{\rm eff}$
and the top horizontal axis shows the ratio $\tau_{xy}/T_{\rm dur}$.}
\label{fig_sigma}
\end{figure}

\subsubsection{The Case of $n\rightarrow\infty$}
In general, the values of $\sigma_z^2$ and $n_{\rm eff}$ need to be calculated numerically.
However, for $n\rightarrow\infty$ and with some specific light-curve variation models,
$\sigma_z^2$ can be derived analytically. Let's take the damped random walk model
as an example. Its covariance function is given by
\begin{equation}
\gamma(k) = \sigma_{\rm d}^2\exp\left(-\frac{|k|\Delta t}{\tau_d}\right),
\end{equation}
and its auto-correlation function is given by
\begin{equation}\label{eqn_rho}
\rho(k) = \frac{\gamma(k)}{\gamma(0)}= \exp\left(-\frac{|k|\Delta t}{\tau_d}\right),
\end{equation}
where $\sigma_{\rm d}$ is the long-term standard deviation and $\tau_{\rm d}$ is the damping timescale
for the variations to decay back to the mean. Assume that $x(t)$ and $y(t)$ follows
the damped random walk model with a damping timescale of $\tau_{x}$ and $\tau_y$, respectively.
In the case of $n\rightarrow\infty$, the factor $(1-k/n)$
can be neglected and Equation~(\ref{eqn_norm}) is simplified into
\begin{equation}\label{eqn_std1}
\sigma_z^2 =\frac{1}{n}\left[ 1 + 2 \left(e^{\Delta t/\tau_{xy}}-1\right)^{-1}\right]
=\frac{1}{n}\coth\left(\frac{\Delta t}{2\tau_{xy}}\right),
\end{equation}
where
\begin{equation}\label{eqn_tauxy}
\tau_{xy} = \left(\frac{1}{\tau_x} + \frac{1}{\tau_y} \right)^{-1}.
\end{equation}
Equation~(\ref{eqn_std1}) illustrates that when the sampling interval $\Delta t$ is much larger
that the damping timescale $\tau_{xy}$, the light curves can be simply treated as white noise and
the variance $\sigma_z^{2} \approx 1/n$. In the opposite case, the auto correlation of the light curves
cannot be neglected and it effectively increases the variance. As a result, there will be a higher
chance of observing a large cross-correlation coefficient.

In practice, each measurement is associated with an uncertainty. The auto-correlation function
in Equation~(\ref{eqn_rho}) is then written as
\begin{equation}\label{eqn_rho_err}
\rho(k) =\left(1+\frac{\epsilon^2}{\sigma_{\rm d}^2}\right)^{-1} \left[\exp\left(-\frac{|k|\Delta t|}{\tau_d}\right) + \frac{\epsilon^2}{\sigma_{\rm d}^2}\delta(k\Delta t)\right],
\end{equation}
where $\delta(x)$ is the Dirac's delta function and $\epsilon$ is the measurement error.
As a result, the variance of $z$ in Equation~(\ref{eqn_std1}) is written as
\begin{eqnarray}\label{eqn_std2}
\sigma_z^2 &=&\frac{1}{n}\left[ 1 + 2 \left(e^{\Delta t/\tau_{xy}}-1\right)^{-1}\right.\nonumber\\
&&\left.\qquad\times\left(1+\frac{\epsilon_x^2}{\sigma_{x}^2}\right)^{-1}\left(1+\frac{\epsilon_y^2}{\sigma_{y}^2}\right)^{-1}\right],
\end{eqnarray}
where $\epsilon_x$ and $\epsilon_y$ are the measurement errors of $x$ and $y$, respectively.
We can find that measurement errors tend to reduce the cross-correlation coefficient.
If either one of the two light curves is dominated by random errors ($\epsilon_x\gg\sigma_x$ or $\epsilon_y\gg\sigma_y$),
the light curve can be treated as white noises and there will be $\sigma_z^2=1/n$.

For $\tau_{xy}\gg\Delta t$, there is an approximate expression for $\sigma_z$ as
\begin{equation}\label{eqn_std3}
\sigma_z^2 \approx \frac{2\tau_{xy}}{n\Delta t}\left(1+\frac{\epsilon_x^2}{\sigma_{x}^2}\right)^{-1}\left(1+\frac{\epsilon_y^2}{\sigma_{y}^2}\right)^{-1},
\end{equation}
where we note that $n\Delta t$ is equal to the duration of the light curves. If neglecting the measurement errors,
the effective number can be simply estimated by $n_{\rm eff}=n\Delta t/2\tau_{xy}$.
When $\tau_{xy}$ is even larger than the time duration of light curves, the auto-correlation function (see
Equation~\ref{eqn_rho}) can be approximated as a constant ($\rho(k)\approx1$). As a result, the variance $\sigma_z^2$ approaches
unity regardless of the value of $\tau_{xy}$.
In Figure~\ref{fig_sigma}, we plot $\sigma_z^2$ versus $\tau_{xy}$
for different sampling intervals, where $\sigma_z^2$ is calculated using Equations~(\ref{eqn_norm}) and (\ref{eqn_rho}) and
the time duration of light curves is fixed to 200 days. As can be seen, the estimated $\sigma_z^2$ is independent
of the sampling interval when $\tau_{xy}\gg \Delta t$.

It should be emphasized that the above equations are valid solely for stationary light curves, in which both
their means and covariances are time-independent. Previous studies have shown that AGN light curves can be generally described by
the damped random walk model, which is a stationary Gaussian process (e.g., \citealt{Kelly2009}).
Nevertheless, in some cases AGN light curves exhibit prominent long-term trends. A common
procedure when calculating the cross-correlation function is detrending the light curves using a low-order polynomial so as to
suppress long-term non-stationary component (e.g., \citealt{Li2013, Peterson2014, F92020}).
For a damped random walk process, when the duration of a light curve is shorter than the damping timescale,
the estimated mean and covariance are no longer constant, but instead depend on the used time segment.
As a result, the estimations of $\sigma_z$ and $n_{\rm eff}$ with Equations~(\ref{eqn_norm}) and (\ref{eqn_neff1})
are no longer valid. Nevertheless, we stress that the distribution $p(z)$ still asymptotically approaches a normal distribution.

\begin{figure}[!t]
\centering
\includegraphics[width=0.45\textwidth]{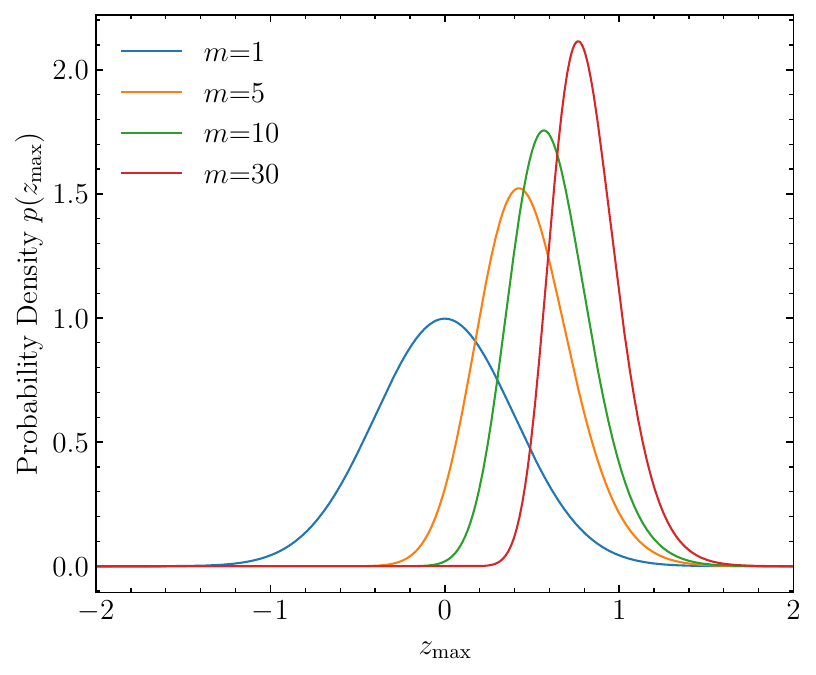}
\includegraphics[width=0.45\textwidth]{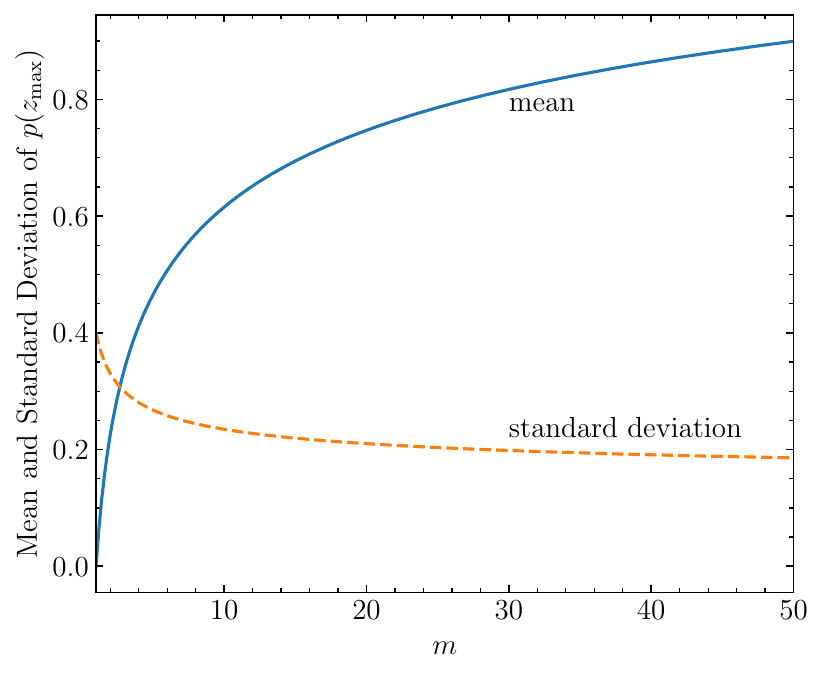}
\caption{(Top) The probability density distribution of $p(z_{\rm max})$ for different $m$ (see Equation~\ref{eqn_prmax2}).
(Bottom) The mean and standard deviation of $p(z_{\rm max})$ with $m$. The value of $\sigma_z$ is set
to 0.4. Here, $m$ can be regarded as the number of independent time-lag bins used to search for $z_{\rm max}$ when calculating the ICCF.}
\label{fig_prmax}
\end{figure}

\begin{figure*}[th!]
\centering
\includegraphics[width=0.8\textwidth]{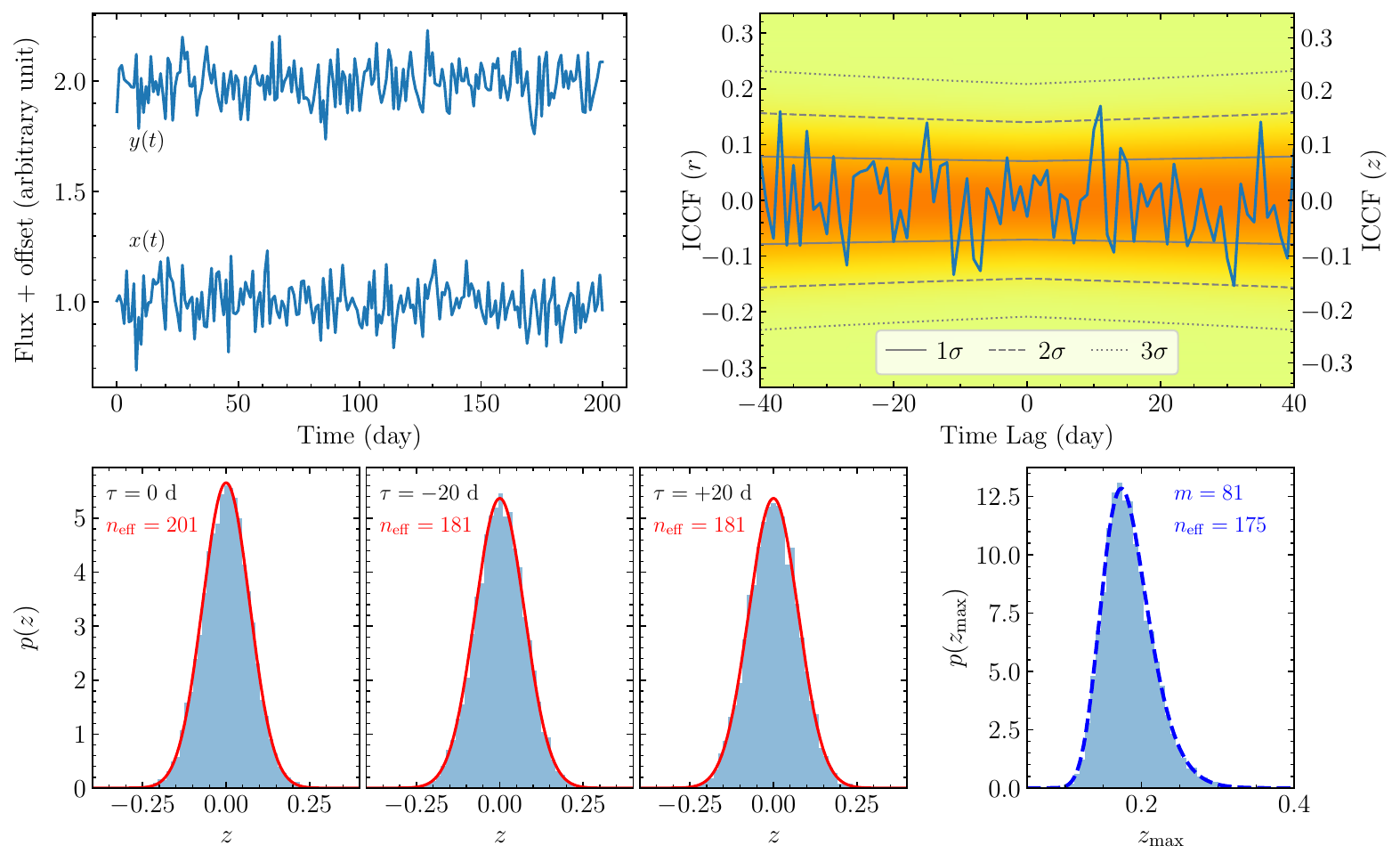}
\caption{ The top left panel shows an example of mock white-noise light curves.
The top right panel shows the ICCF over a time lag range of $(-40, +40)$ days.
The underlying colored heat map shows the probability distribution of $r$ or $z$. The corresponding 1$\sigma$, 2$\sigma$, and 3$\sigma$
confidence regions are highlighted by grey solid, dashed, and dotted lines, respectively.
The bottom left three panels show the histogram distributions of $z$ at $\tau=0$, $-20$, and $+20$ days from $10^4$ pairs of mock light curves, respectively. The red solid lines represent the normal distribution with a variance of $\sigma_z^2=1/n_{\rm eff}$, which is estimated by Equation~(\ref{eqn_neff1}). The bottom rightmost panel shows the histogram distribution of $z_{\rm max}$ searched over a range of time lags $(-40, +40)$ days with a bin width of one day. The red dashed line represents the best-fit distribution using Equation~(\ref{eqn_prmax2})
with $\sigma_z^2=0.0057$ (or equivalently $n_{\rm eff}=175$) and $m=81$.
\label{fig_pz_noise}}
\end{figure*}

\subsection{The Peak Cross-correlation Coefficient}
The peak cross-correlation coefficient is determined by searching for the peak value
of the cross-correlation function over a specific range of time lags (\citealt{Gaskell1987}). Since each
cross-correlation coefficient follows a normal distribution, and assuming the null hypothesis that these normal distributions
are independent, the cumulative distribution of the peak cross-correlation coefficient can be written as
\begin{equation}\label{eqn_Prmax}
P(z_{\rm max}) \propto \prod_{i=1}^{m} G\left(\frac{z_{\rm max}}{\sigma_{z,i}}\right),
\end{equation}
where $m$ is the number of independent time-lag bins searched over, $\sigma_{z, i}$ is the standard deviation
at $i$-th bin, and $G(x)$ is the cumulative normal distribution given by
\begin{eqnarray}
G\left(x\right) &= & \frac{1}{\sqrt{2\pi}}\int_{-\infty}^{x} e^{-y^2/2} dy = \frac{1}{2}{\rm erfc}\left(-\frac{x}{\sqrt{2}}\right),
\end{eqnarray}
where ${\rm erfc}(x)$ is the complementary error function. Differentiating Equation~(\ref{eqn_Prmax})
with respect to $z_{\rm max}$ and assuming that all $\sigma_j$ have the same value $\sigma_z$,
the probability distribution is written as
% \begin{eqnarray}\label{eqn_prmax}
% p(z_{\rm max}) &=& \frac{dP(z_{\rm max})}{dz_{\rm max}} \nonumber\\
% &\propto& \sum_{j=1}^{m} \frac{1}{\sigma_{z, j}}g\left(\frac{z_{\rm max}}{\sigma_{z, j}}\right)
% \prod_{i=1, i\neq j}^m G\left(\frac{z_{\rm max}}{\sigma_{z,j}}\right),
% \end{eqnarray}
%where $g(x)$ is Gaussian function.
%If all $\sigma_j$ have the same value $\sigma_z$, $p(z_{\rm max})$ can be simplified into
\begin{eqnarray}\label{eqn_prmax2}
p(z_{\rm max})& =&\frac{dP(z_{\rm max})}{dz_{\rm max}} \nonumber \\
&\propto& \frac{m}{\sigma_z}g\left(\frac{z_{\rm max}}{\sigma_z}\right)
\left[G\left(\frac{z_{\rm max}}{\sigma_z}\right)\right]^{m-1},
\end{eqnarray}
where $g(x)$ is the Gaussian function.

In the above derivation, we assume that the normal distribution in each time-lag bin
is independent. In practice, this is not true as the cross-correlation coefficients among adjacent
time-lag bins are indeed correlated owing to the auto-correlations in light curves.
This effectively reduces the value $m$ in Equation~(\ref{eqn_prmax2}).
Moreover, for a different time-lag bin, the corresponding standard deviation $\sigma_{z, j}$ is typically
different because both the number of data points and the effective duration of light curves
change with time lag. However, we will show below that Equation~(\ref{eqn_prmax2}) still provides
a reasonable approximation to the probability distribution of $z_{\rm max}$.

In the top panel of Figure~\ref{fig_prmax}, we plot the probability density distribution $p(z_{\rm max})$ in Equation~(\ref{eqn_prmax2})
for different $m$ by setting $\sigma_z=0.4$. As expected, the probability of observing a large $z_{\rm max}$ increases with $m$.
This is because there is a higher chance of randomly observing a large $z_{\rm max}$ when searching
over more time-lag bins. This is also reflected in the bottom panel of Figure~\ref{fig_prmax},
where the mean of $z_{\rm max}$ increases with $m$.  Notably, the standard deviation of
$z_{\rm max}$ is quite insensitive to $m$ and its value tends to be roughly one half of $\sigma_z$ for large $m$.

% In Equation~(\ref{eqn_prmax2}), the value of $\sigma_z$ can be estimated using Equation~(\ref{eqn_norm}).
% It is not straightforward to dervie an analytical expression for $m$, however, we find that
% \begin{equation}\label{eqn_m}
%  m \approx \frac{\tau_{\rm end}-\tau_{\rm beg}}{2\tau_{xy}},
% \end{equation}
% provides a rough approximation, if the light curves follow the damped random walk model.
% Here, $\tau_{\rm beg}$ and $\tau_{\rm end}$ represent the begining and ending time lags of the range
% used to search for $z_{\rm max}$.

We stress that the probability $p(z_{\rm max})$ has appropriately included the ``look-elsewhere'' effect.
Because $z_{\rm max}$ is obtained by searching over a number of time-lag bins, one should not simply calculate
the significance level using $p(z)$ at the time-lag bin corresponding to $z_{\rm max}$, which will clearly lead to
an overestimated significance level of $z_{\rm max}$.

\subsection{Calculating the Cross-correlation Function}
The realistic light curves in AGN monitoring are usually irregularly sampled
and their observing cadences are also not contemporaneous.
The ICCF method is widely adopted to calculate cross-correlation coefficients (\citealt{Gaskell1987}).
In this approach, two rounds of cross-correlation coefficients are computed, with only one light curve being interpolated in each round.
The final cross-correlation coefficient is then assigned as the average of the two rounds.
Specifically, given a time lag $\tau$, firstly shift $x(t)$ in time with $\tau$ and extract the segment in $x(t+\tau)$
overlapping with $y(t)$, which we denote $x'(t)$. Then interpolate $y(t)$ onto the time points same as $x'(t)$.
As such, we obtain two light curves $x'(t)$ and $y'(t)$ with the same duration and sampling rate.
We then directly compute their cross-correlation coefficient as below.
The second round repeats the same procedure but shifts $y(t)$ in time with $-\tau$ and implements interpolation on $x(t)$.

After interpolation, the cross-correlation coefficient of the light-curve pairs $x'(t)$ and $y'(t)$ in
Equation~(\ref{eqn_rtau}) is then rewritten into
\begin{equation}\label{eqn_r}
r = \frac{\sum_{i=1}^{n} (x'_i-\bar x')(y'_i-\bar y')}{\sqrt{\sum_{i=1}^n(x'_i-\bar x')^2\sum_{i=1}^{n}(y'_i-\bar y')^2}},
\end{equation}
where $n$ is the number of points in the light curves,
\begin{equation}
\bar x' = \frac{1}{n}\sum_{i=1}^{n}x'_i,
\end{equation}
and
\begin{equation}
\bar y' = \frac{1}{n}\sum_{i=1}^{n}y'_i.
\end{equation}
The linear interpolation is used throughout our calculations.

\begin{figure*}[th!]
\centering
\includegraphics[width=0.8\textwidth]{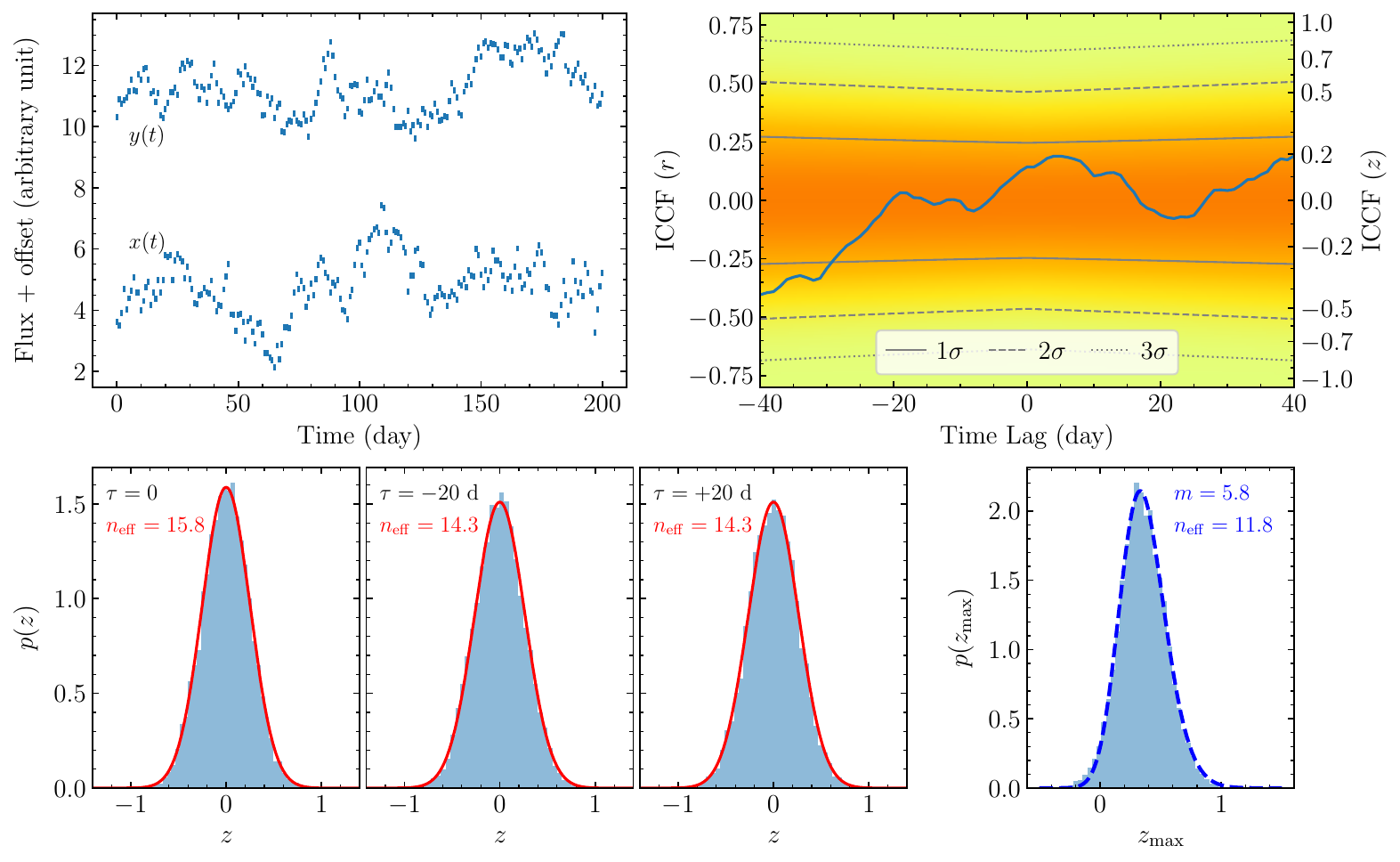}
\caption{Same as Figure~\ref{fig_pz_noise}, but for regularly sampled AGN light curves.}
\label{fig_pz_sim}
\end{figure*}

\section{Simulation Tests}\label{sec_tests}

\subsection{White-noise Light Curves}
For the sake of illustration, we simply generate white-noise light curves with a total duration of 200 days and a daily
cadence, i.e., 201 epochs. We then compute the ICCF using Equation~(\ref{eqn_r}) and determine
the peak cross-correlation coefficient on a time-lag
range of $(-40, +40)$ days with a bin width of one day, which is equal to the sampling interval of the light curves.

In the top panels of Figure~{\ref{fig_pz_noise}}, we show an example of generated light curves and the resulting ICCF.
In the bottom panels, we plot the histograms of the cross-correlation coefficient at $\tau=0$, $-20$, and $+20$ days, and
the histogram of $z_{\rm max}$ from $10^4$ pairs of mock light curves. As expected, the cross-correlation coefficients
follow normal distributions (red line) quite well, where the variances are proportional to the reciprocal of the
number of data points.
The distribution of $z_{\rm max}$ can be well-fitted using
Equation~(\ref{eqn_prmax2}), with $\sigma_z^2=0.0057\approx1/175$ and $m=81$. Since the cross-correlation coefficient
in each time-lag bin is independent, the value of $m$ is exactly equal to the number of time-lag bins.
The number of points involved in calculating
the cross-correlation coefficient changes with the input time lag, therefore, the value of $\sigma_z$
in $p(z_{\rm max})$ is slightly different from that of $p(z)$ at $\tau=0$. Using the above best-fit distribution $p(z_{\rm max})$,
the probability of $z_{\rm max}>0.3$ (corresponding to $r>0.29$) and $0.4$ (corresponding to $r>0.38$) are 0.003 and $5\times10^{-6}$, respectively.

For white-noise data, there is indeed an exact analytic expression for the probability distribution of $r$, as first
found by \cite{Fisher1921}
\begin{equation}
p(r) = \frac{\Gamma[(n-1)/2]}{\sqrt{\pi}\Gamma[(n-2)/2]}(1-r^2)^{(n-4)/2},
\end{equation}
where $\Gamma[x]$ is the gamma function. It is easy to verify that for large $n$, $p(r)\propto e^{(n-4)\ln(1-r^2)/2}
\approx e^{-(n-4)r^2/2}$. Therefore, $p(r)$ (and $p(z)$)
approaches the normal distribution with a variance of $1/(n-4)\approx 1/n$.

\begin{figure*}[th!]
\centering
\includegraphics[width=0.8\textwidth]{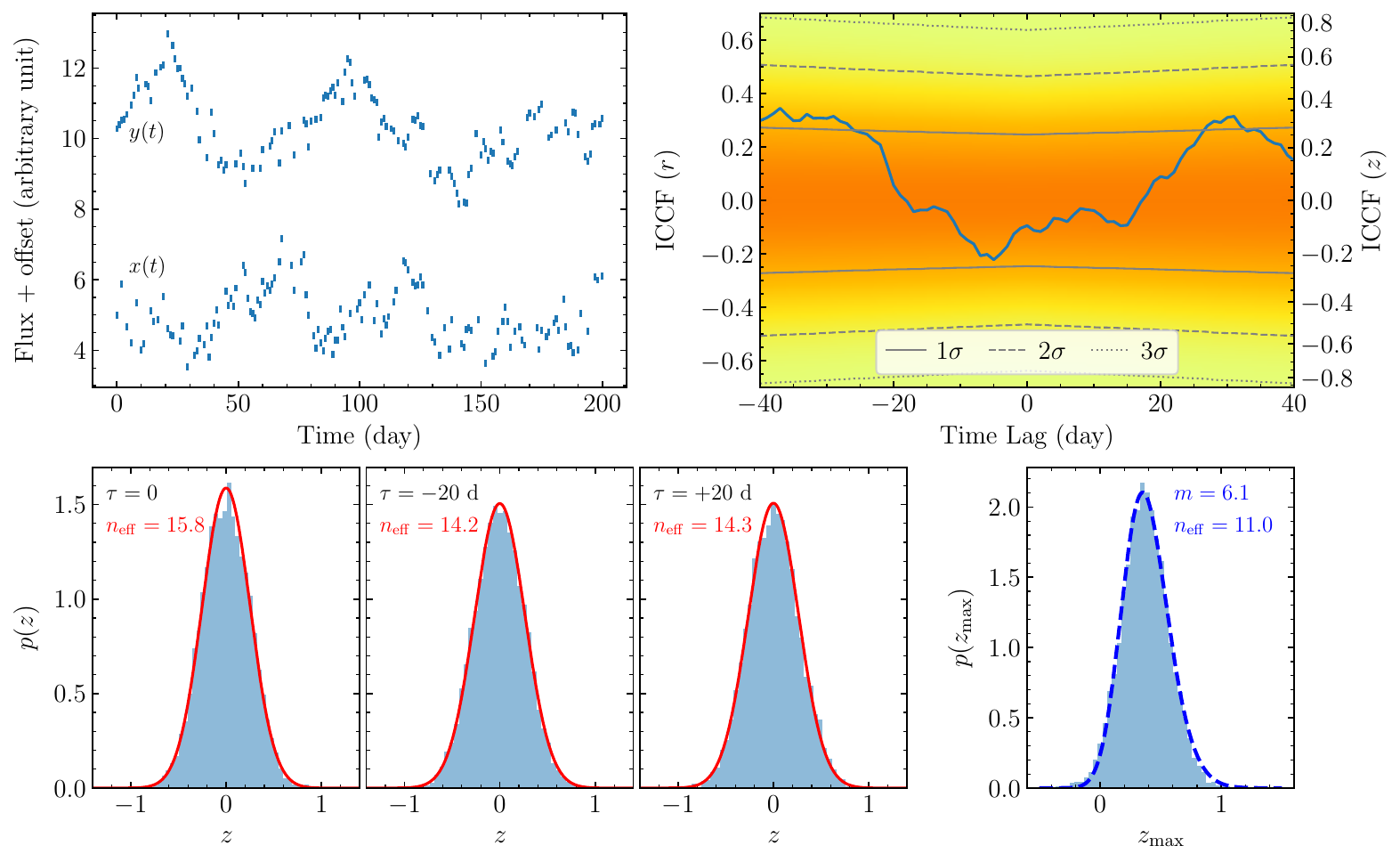}
\caption{Same as Figure~\ref{fig_pz_sim}, but for irregularly sampled AGN light curves. The sampling pattern differs between
the two light curves.}
\label{fig_pz_sim2}
\end{figure*}

\begin{figure*}[th!]
\centering
\includegraphics[width=0.8\textwidth]{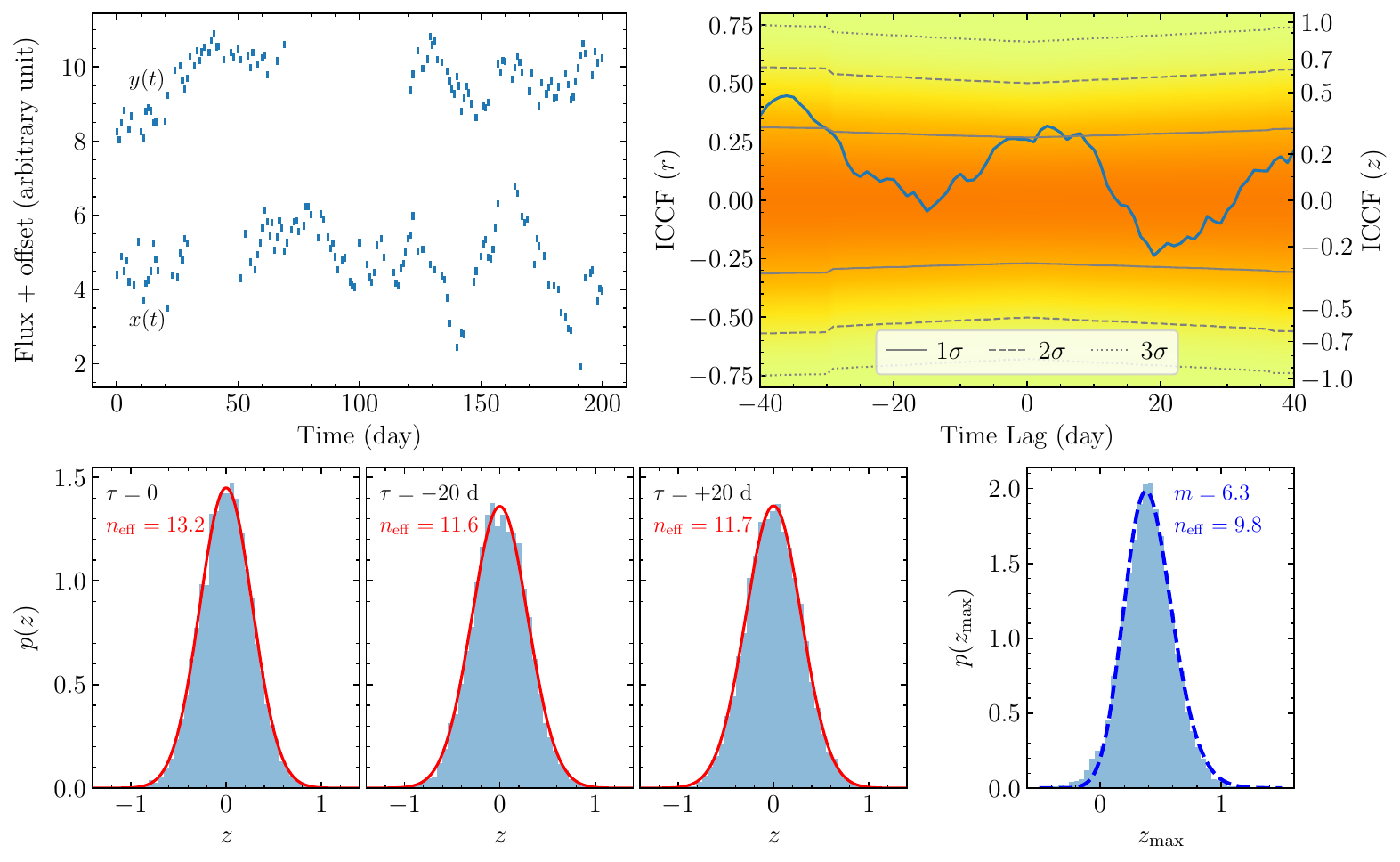}
\caption{Same as Figure~\ref{fig_pz_sim}, but for irregularly sampled AGN light curves with gaps.}
\label{fig_pz_sim3}
\end{figure*}

\begin{figure*}[th!]
\centering
\includegraphics[width=0.8\textwidth]{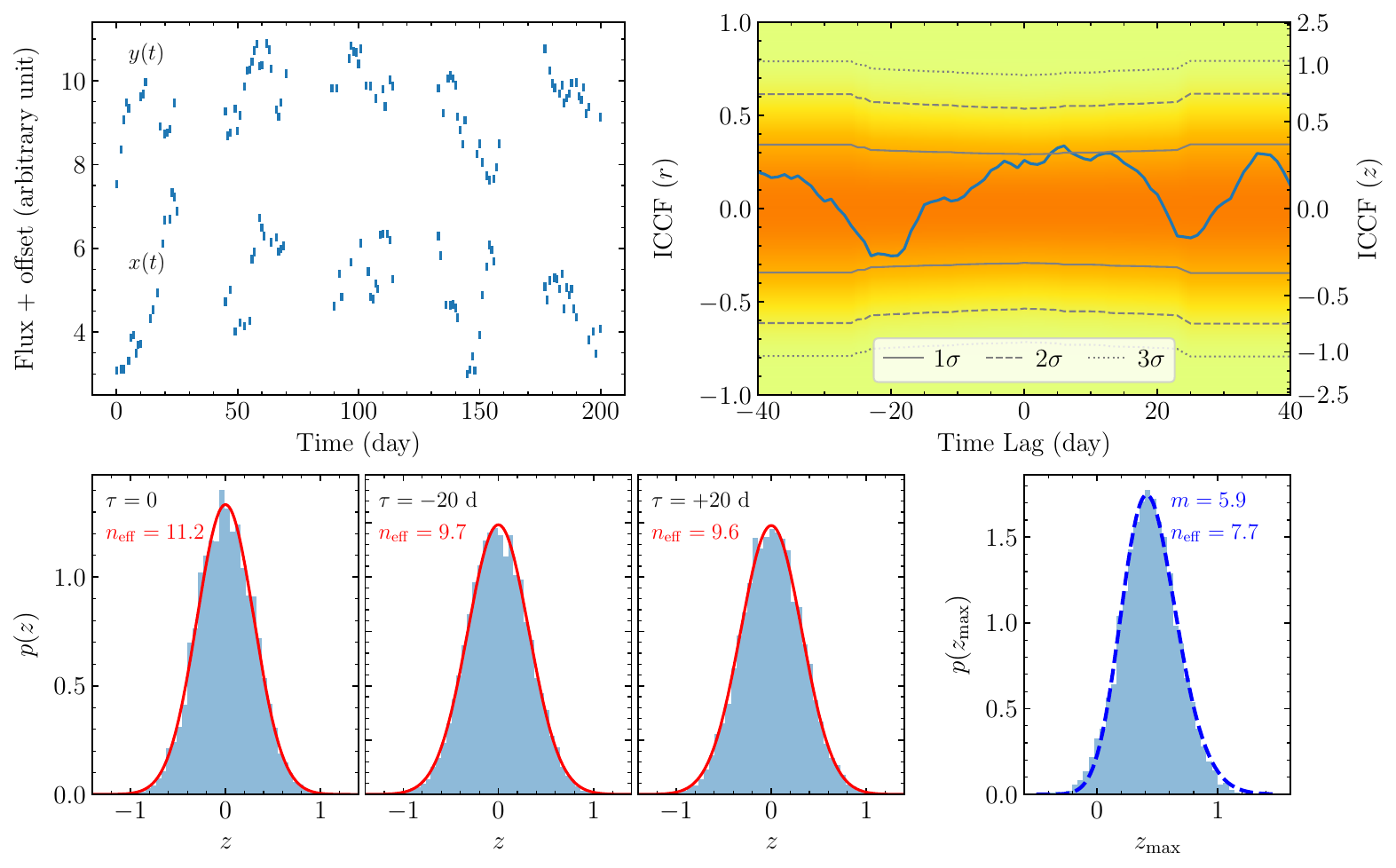}
\caption{Same as Figure~\ref{fig_pz_sim}, but for irregularly sampled AGN light curves with periodical gaps.
The gaps occur at the same epochs in both light curves, with a duty cycle comparable to that of the seasonal gaps ($\sim$40\%).}
\label{fig_pz_sim3_same}
\end{figure*}

\subsection{AGN Light Curves: Regularly Sampled}\label{sec_agn_reg}
We use two independent damped random walk models to generate mock AGN light curve pairs. The one model with parameters $\sigma_{x}=1.0$ (arbitrary unit) and $\tau_{x}=10$ days, and the other model with parameters $\sigma_{y}=1.0$ (arbitrary unit) and $\tau_{y}=20$ days, yielding $\tau_{xy}=6.7$ days in Equation~(\ref{eqn_tauxy}). These damping timescales are chosen to be much shorter than the time duration of light curves, ensuring the resulting means
and covariances in Equation~(\ref{eqn_r}) can be treated as time-independent.
The error of each point in both light curves is assigned $\epsilon=0.1$ (arbitrary unit).
Here, the units of $\sigma_{\rm d}$ and $\epsilon$ are irrelevant since they cancel out when calculating the cross-correlation coefficients.
The time duration is 200 days and the sampling cadence is daily. This leads to $n_{\rm eff}=15.8$ from Equations~(\ref{eqn_neff1})
and (\ref{eqn_rho_err}). Again, the cross-correlation function is computed over a time-lag range of ($-40$, $+40$) days
with a bin width of one day.

The top two panels of Figure~\ref{fig_pz_sim} showcase a pair of mock light curves and the ICCF, respectively. The bottom left three panels plot the histograms of $z$ at $\tau=0$, $-20$, and $+20$ days from $10^4$ pairs of mock light curves, respectively.  The rightmost panel plots the histogram of $z_{\rm max}$.
Compared with Figure~\ref{fig_pz_noise}, the histograms of both $z$ and $z_{\rm max}$ span a significantly broader range,
reflecting the effects of auto-correlations in the light curves.

The theoretical normal distribution $p(z)$ with $\sigma_z^2=1/n_{\rm eff}$ (red line) remarkably matches the histogram of $z$,
indicating the validity of the analytical expressions in Section~\ref{sec_ccf}.
The best-fit distribution $p(z_{\rm max})$ gives $\sigma_z^2=0.085\approx1/11.8$ and $m=5.8$.
This small value of $m$ renders the long tail of $p(z_{\rm max})$ less prominent.
We note that, unlike to the white-noise case, the value of $m$ is far smaller than the number of time-lag bins (=81)
used to search for $z_{\rm max}$. As described above, this is because the cross-correlation coefficients between adjacent time-lag bins
are no longer independent, but indeed correlated. The effective number of time-lag bins is thereby significantly reduced.
We find that the value of $m$ can be roughly estimated by $(\tau_{\rm end}-\tau_{\rm beg})/2\tau_{xy}$, where $\tau_{\rm beg}$ and
$\tau_{\rm end}$ are the beginning and ending time lags used to calculate the ICCF. In the present case, the yielded value of
$m\approx6$, close to the above best-fit value.

\begin{figure}[t!]
\centering
\includegraphics[width=0.48\textwidth]{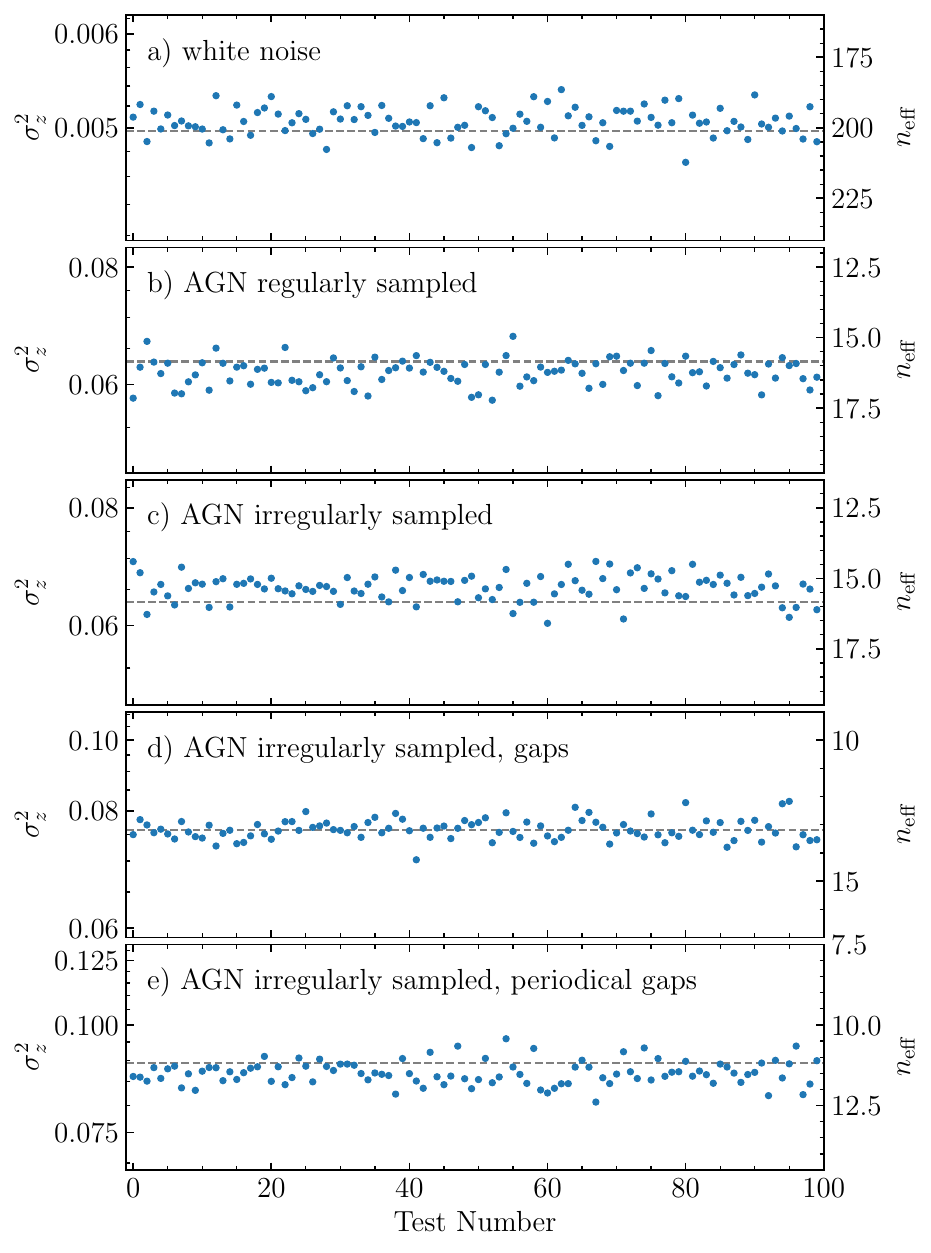}
\caption{The values of $\sigma_z$ and $n_{\rm eff}$ obtained by directly
fitting the histograms of $z$ at $\tau=0$ from simulations using a normal distribution
for a) white-noise light curves, b) AGN regularly sampled light curves, c) AGN irregularly sampled light curves,
d) AGN irregularly sampled light curves with gaps, and e) AGN irregularly sampled light curves with periodical gaps.
The corresponding examples of light curves for these cases are shown in Figures~\ref{fig_pz_noise}-\ref{fig_pz_sim3_same}, respectively.
In the top panel, the grey horizontal dashed line represents $n_{\rm eff}=201$.
In the other panels, the grey horizontal dashed line represents
the theoretical estimates of $\sigma_z$ (or equivalent $n_{\rm eff}$)  using Equation~(\ref{eqn_neff1})
and the input $\tau_x$ and $\tau_y$.}
\label{fig_sim_ndeff}
\end{figure}

\begin{figure*}[th!]
\centering
\includegraphics[width=0.8\textwidth]{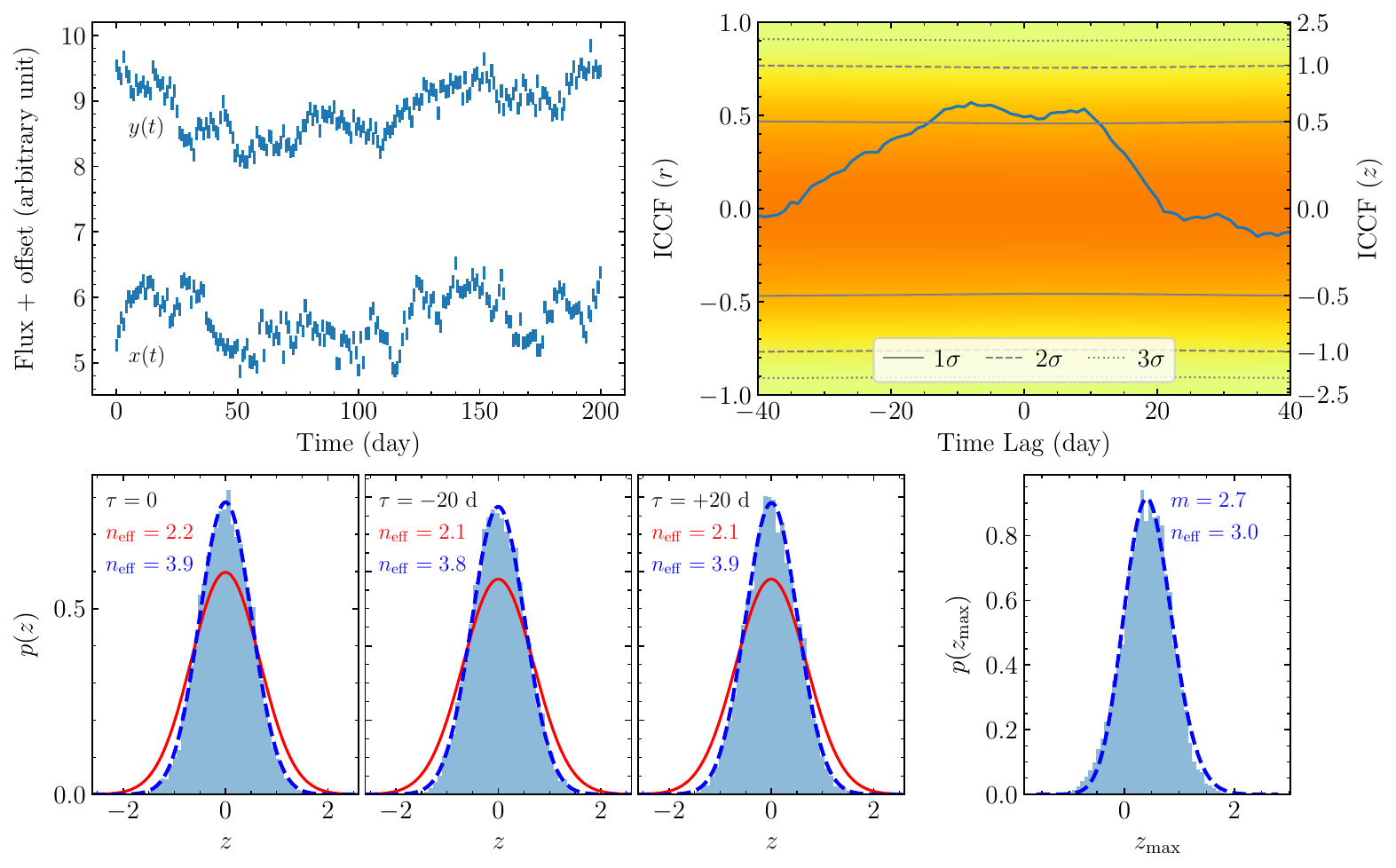}
\caption{Same as Figure~\ref{fig_pz_noise}, but for regularly sampled AGN light curves with the damping timescales
of the damped random walk models set to be $\tau_x=100$ days and $\tau_y=200$ days in the simulations.
In the top right panel, the colored heat map shows the probability distribution of $r$ or $z$, which is calculated using
the best-fit variances from mock light curves.
In the bottom panels, the red numbers
represent theoretical estimates using the input damping timescales, while the blue numbers represent
the best-fit values to the histograms from simulations.}
\label{fig_pz_sim_timescale}
\end{figure*}

\subsection{AGN Light Curves: Irregularly Sampled}
As mentioned above, the formulae for the probability distributions derived in Section~\ref{sec_ccf} require
regularly sampling. This condition is generally not satisfied in realistic observations.
For irregularly sampled data, as an approximation, the sampling interval
can be assigned by the mean sampling interval as
\begin{equation}
\Delta t = \frac{T_{\rm dur}}{n-1},
\end{equation}
where $T_{\rm dur}$ is the duration of the light curves and $n$ is the number of points.
Here, all the quantities refer to the interpolated light curves, therefore, in the two rounds of
calculations for the cross-correlation coefficients, the duration $T_{\rm dur}$ is identical. Only
the number of points $n$ is different between the two rounds, depending on which light curve is being interpolated.
However, Equations~(\ref{eqn_std3}) illustrates that the variance of $z$ depends on the ratio
$2\tau_{xy}/n\Delta t\approx 2\tau_{xy}/T_{\rm dur}$. This implies that the variances of $z$ in the two
rounds of calculations are similar and
the averaged $z$ of the two rounds possesses the same variance. This also means that the variance $\sigma_z$
and effective number $n_{\rm eff}$ are not sensitively
affected by the sampling rates of light curves.

We use the same procedure as in the preceding section to generate mock light curves. To mimic irregular sampling,
we randomly discard one third of the points from both light curves. The set of discarded points differs between the two light
curves, however, the resulting $\sigma_z$ and $n_{\rm eff}$ are insensitive to the light-curve sampling patterns.
%This results in $n_{\rm eff}=15.8$ at $\tau=0$ from Equations~(\ref{eqn_neff1}) and (\ref{eqn_rho_err}) using the input $\tau_x$ and $\tau_y$.
We then repeat this procedure to generate $10^4$ light curve pairs. Note that the sampling patterns are kept identical for all
light curve pairs. The top panels of Figure~\ref{fig_pz_sim2} shows an example of a mock light curve pair and
the resulting ICCF, respectively. In the bottom panels of Figure~\ref{fig_pz_sim2}, we plot the
histograms of $z$ at $\tau=0$, $-20$, and $+20$ days, as well as the histogram of $z_{\rm max}$.
Again, we can find that the histograms of $z$ are in good agreement with the expected normal distributions (red lines),
where the variances are estimated from Equations~(\ref{eqn_neff1}) and (\ref{eqn_rho_err}) using the input
$\tau_x$ and $\tau_y$.
The histogram of $z_{\rm max}$ is fitted using Equation~(\ref{eqn_prmax2}) with $\sigma_z^2=0.09\approx1/11.0$
and $m=6.1$. These values are generally similar to the case of regular sampling in Section~\ref{sec_agn_reg}.

In Figure~\ref{fig_pz_sim3}, we perform another experiment test by inserting sampling gaps in the light curves.
We set a gap of $\sim$20 days between the interval 30-50 days for the light curve $x(t)$ and a gap of $\sim$50 days
between 70-120 days for the light curve $y(t)$. By taking into account gaps, the sampling interval should
be estimated by
\begin{equation}\label{eqn_dt}
\Delta t = \frac{T_{\rm dur}-T_{\rm gap}}{n-1-n_{\rm gap}},
\end{equation}
where $n_{\rm gap}$ is the number of gaps and $T_{\rm gap}$ is their total duration. A gap is identified
when its duration is substantially longer than the typical sampling interval. The term $n_{\rm gap}$ in the denominator of
Equation~(\ref{eqn_dt}) arises because each gap renders one data point useless when computing the
sampling interval.
Since the gap durations are different between the two light curves,
the estimated $\sigma_z$ and $n_{\rm eff}$ are different accordingly in the two rounds of calculations,
again depending on which light curve is being interpolated.
We simply assign the final $\sigma_z^2$ as the average of $\sigma_z^2$ in the two rounds of calculations (note that $n_{\rm eff}=1/\sigma_z^2$).
The bottom panels of Figure~\ref{fig_pz_sim3} illustrate that such an approximate estimate is in good agreement with
the simulation results. The probability distribution $p(z_{\rm max})$ in the bottom right panel of
Figure~\ref{fig_pz_sim3} is well fitted with $\sigma_z^2=0.102\approx1/9.8$ and $m=6.3$.

For multi-year monitoring campaigns, light curves typically exhibit seasonal gaps. To mimic this effect,
we insert periodical gaps into the mock light curves, with a duty cycle comparable to that of the seasonal gaps ($\sim$40\%).
Figure~\ref{fig_pz_sim3_same} shows an example of randomly generated light curves.
Compared with Figure~\ref{fig_pz_sim3}, the $n_{\rm eff}$ values for $p(z)$ at $\tau=0$, $-20$, and $+20$ days are overall smaller
owing to the reduced number of data points. By contrast, the $m$ values for $p(z_{\rm max})$ are similar, both around $m\sim6$, which
is consistent with the approximation $(\tau_{\rm end}-\tau_{\rm beg})/2\tau_{xy}$ mentioned in Section~\ref{sec_agn_reg}.

In Figures~\ref{fig_pz_noise}-\ref{fig_pz_sim3_same}, for comparison, we plot the normal distributions with the
theoretically estimated variances using the input $\tau_x$ and $\tau_y$.
From simulations, we can also directly fit the obtained histograms of $z$ with a normal
distribution. Owing to statistical fluctuations, we expect that the best-fit $\sigma_z^2$ or $n_{\rm eff}$
exhibits a scatter around the theoretical value. In Figure~\ref{fig_sim_ndeff}, we show the obtained $\sigma_z^2$ and
$n_{\rm eff}$ from 100 tests for each case of light-curve configurations as described above. Note that for irregularly sampled
AGN light curves, the sampling patterns are random across different tests.

\subsection{AGN Light Curves: the Damping Timescale Comparable to the Time Duration}\label{sec_bias}
As stated above, the expressions for $\sigma_z^2$ and $n_{\rm eff}$ derived in Section~\ref{sec_ccf}
are only valid for stationary light curves. This requires the characteristic timescale of light curves to be
much shorter than the time durations. In the above simulations, we adopt the damping timescales of $\tau_x=10$ and
$\tau_y=20$ days. As a comparison, the time duration of light curves are 200 days.
To investigate the impact of the damping timescale on $\sigma_z^2$,  in Figure~\ref{fig_pz_sim_timescale}, we perform a simulation test
with the damped timescales set to be $\tau_{x}=100$ days and $\tau_{y}=200$ days for the light curve pair, respectively.
The rest simulation configurations are the same as those in Section~\ref{sec_agn_reg}.
As can be seen from  the top panels of Figure~\ref{fig_pz_sim_timescale}, the exemplary mock light curves
exhibit low-frequency variation trends, which lead to a peak cross-correlation coefficient $z_{\rm max}>0.5$.
Compared to Figure~\ref{fig_pz_sim}, the histograms of $z$ span a much broader range. The histograms still follow the normal distribution, however, their variances are typical smaller than
the theoretical estimates using the input $\tau_{x}$ and $\tau_{y}$.
This is because the limited time durations of light curves effectively reduce the damped timescales,
leading to a smaller variance $\sigma_z^2$ than the theoretical estimate.
In Figure~\ref{fig_sigma}, we show a comparison between the theoretical estimates of $\sigma_z^2$ with those from
simulations for different $\tau_{xy}$. The deviation begins to be evident when the ratio $\tau_{xy}/T_{\rm dur}>0.1$
and increases with $\tau_{xy}/T_{\rm dur}$. For $\tau_{xy}=T_{\rm dur}$, the variance $\sigma_z^2$
is smaller by a factor of two than
the theoretical estimate. This implies that when performing null-hypothesis testing, the theoretical estimate
of $\sigma_z^2$ gives a conservative lower bound for the significance level.

\begin{figure*}
\centering
\includegraphics[width=0.8\textwidth]{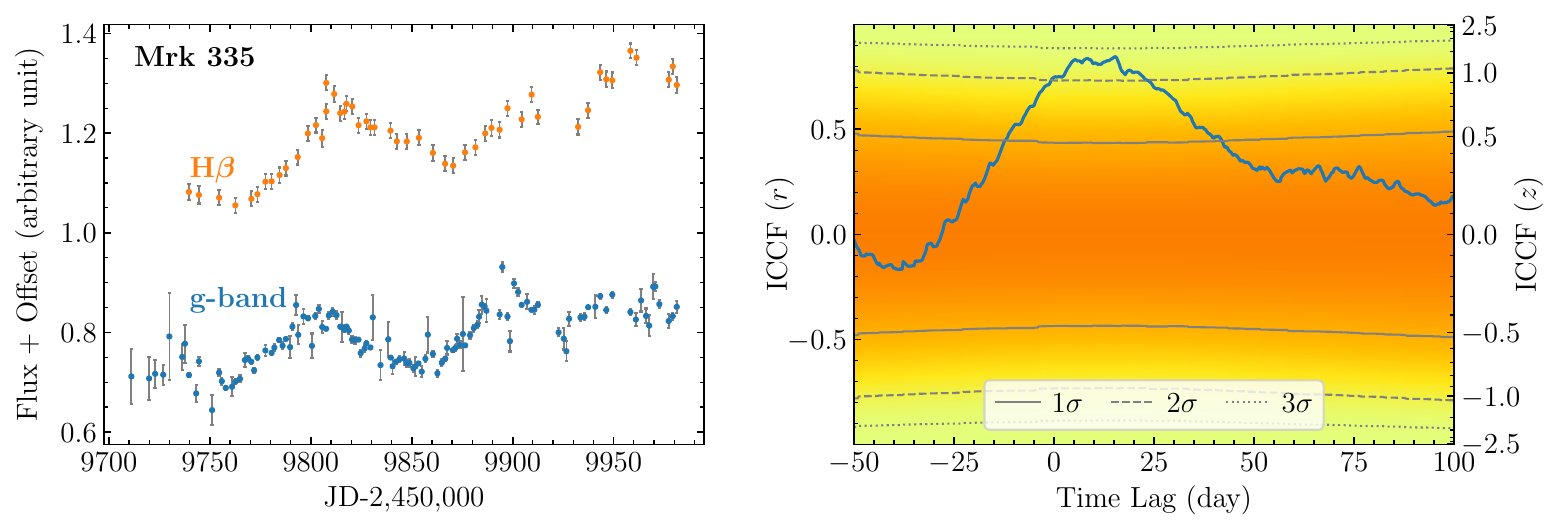}
\includegraphics[width=0.8\textwidth]{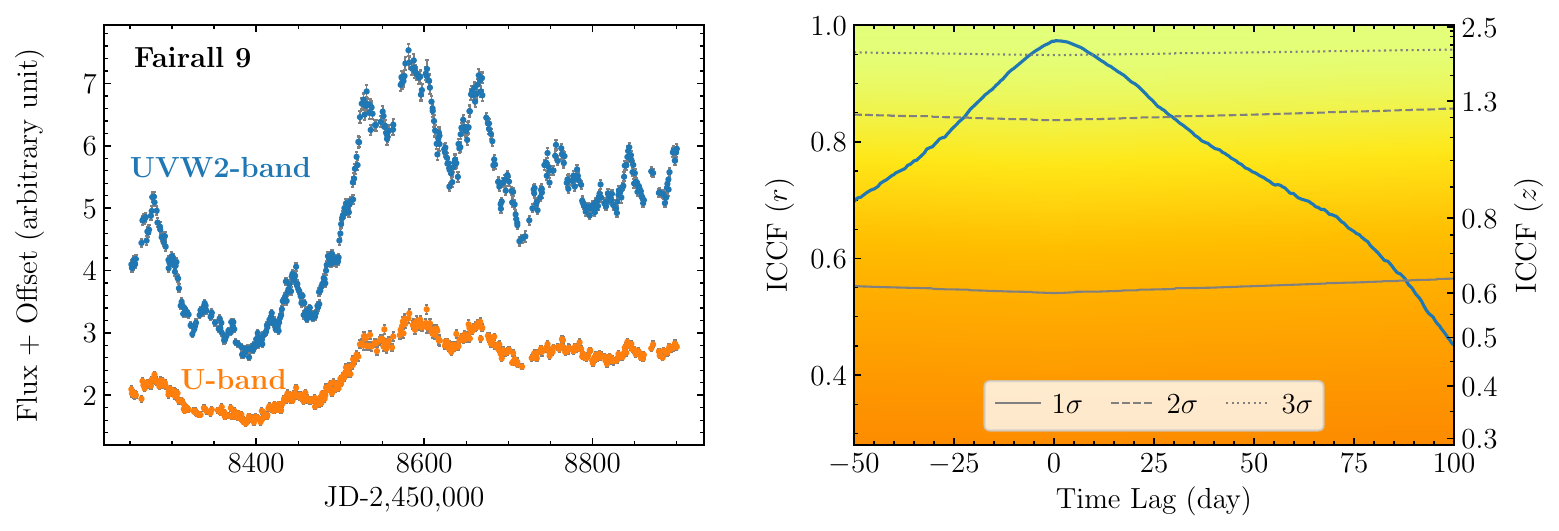}
\includegraphics[width=0.8\textwidth]{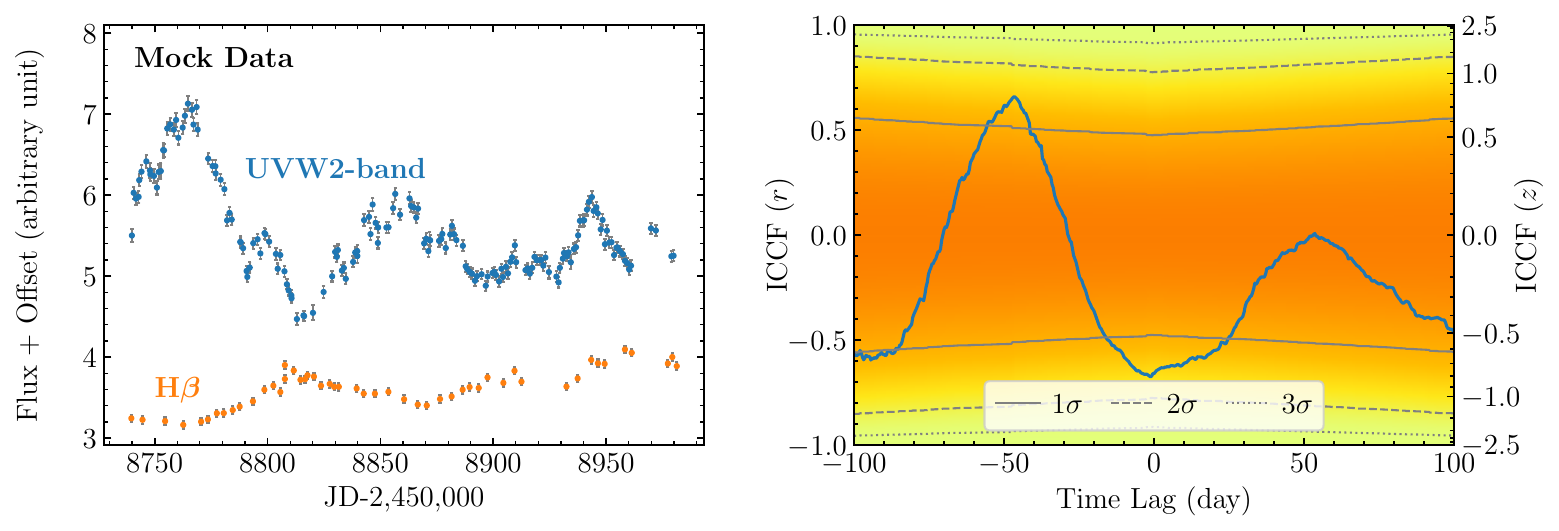}
\caption{Application of null-hypothesis testing to realistic AGN light curves of Mrk 335, Fairall 9, and mock data from top to bottom
panels. The top panels show
the g-band and H$\beta$ light curves of Mrk 335 from \cite{Li2024SARM}; the middle panels show the
{\it Swift} UVW2- and U-band light curves of Fairall 9 form \cite{Edelson2024}; the bottom panels show the mock data
generated by combining the H$\beta$ light curve of Mrk 335 and a segment of the UVW2-band light curve of Fairall 9 (with a time shift).
The right panels show the ICCFs (blue solid lines) and the probability distributions of $r$ or $z$
(colored heat maps). The 1$\sigma$, 2$\sigma$, and 3$\sigma$ confidence regions of the heat map are highlighted by grey solid, dashed, and dotted
lines, respectively.
}
\label{fig_tests}
\end{figure*}

\section{Null-hypothesis Testing for Cross-correlation}\label{sec_null}
As described above, the distribution of $z$ at a given time lag follows a normal distribution for the null
hypothesis of no correlated variations between the two light curves.
This property allows us to directly estimate the significance level of the cross-correlation coefficient
at each time-lag bin. The basic procedure
is as follows. First, fit the two light curves with the damped random walk model and determine the
corresponding best-fit model parameters. Then calculate the ICCF using
Equation~(\ref{eqn_r}) and estimate the associated variances using Equation~(\ref{eqn_norm}).
The determination of the variance at each time lag bin enables a direct calculation of
the probability for a  given cross-correlation coefficient under the null hypothesis, namely,
the light curves are random and uncorrelated.
In the top right panels of Figures~\ref{fig_pz_noise}-\ref{fig_pz_sim_timescale}, we superimpose
a colored heat map to show the probability distribution of $r$ or $z$ at a given time lag, together
with the corresponding 1$\sigma$, 2$\sigma$, and 3$\sigma$ confidence regions.

Regarding the peak cross-correlation coefficient $z_{\rm max}$, we have no analytic expression
for estimating the values of $m$ and $\sigma_z$ involved in the probability distribution $p(z_{\rm max})$
(see Equation~\ref{eqn_prmax2}). We need to employ Monte Carlo simulations to determine
the probability of $z_{\rm max}$ (e.g., \citealt{MaxMoerbeck2014,Li2021, U2022, Wang2024}).
Specifically, we generate a sizable sample of mock light curves that possess
identical properties (such as sampling cadences and signal-to-noise ratios etc) to the
observed light curves. We then use Equation~(\ref{eqn_prmax2})
to fit the obtained distribution of $z_{\rm max}$.
According to simulations presented in the preceding section, as a rough approximation,
the value of $m$ can be estimated by $m\approx (\tau_{\rm end}-\tau_{\rm beg})/2\tau_{xy}$
and the value of $\sigma_z$ can be estimated as the average $\sigma_z$ across all time-lag bins.

To demonstrate the application of the above procedures to realistic AGN light curves, we select two
sets of light curves from previous broad-line and accretion-disk reverberation mapping campaigns in the literature.
The first one consists of the g-band and H$\beta$ light curves of Mrk 335 reported in \cite{Li2024SARM} and
the other one consists of {\it Swift} UVW2- and U-band light curves of Fairall 9 reported in \cite{Edelson2024}.
Figure~\ref{fig_tests} plots these light curves and the corresponding ICCFs.

For Mrk 335, the best-fit damping timescales of the g-band and H$\beta$ light curves are
$\tau_x\sim44$ days and $\tau_y\sim117$ days respectively, resulting in $\tau_{xy}\sim32$ days.
The total time duration of the light curves is 241 days.
The ICCF, computed over a time-lag range from -50 to 100 days, peaks around $\tau\sim10$ days
($r_{\rm max, obs}=0.84$ and $z_{\rm max, obs}=1.24$),
where the cross-correlation coefficients exceed the 2$\sigma$ confidence region ($|z|\lesssim0.99$)
of the null hypothesis. For the probability distribution $p(z_{\rm max})$ given by Equation~(\ref{eqn_prmax2}),
if we simply adopt the parameters $\sigma_z=0.49$ and $m=150/2\tau_{xy}\sim2.3$, the probability for $z_{\rm max}$
larger than the observed $z_{\rm max, obs}$ in the null hypothesis is $\sim0.01$.

For Fairall 9, the best-fit damping timescales of the UVW2- and U-band light curves are $\tau_x\sim291$ days and $\tau_y\sim358$ days respectively, resulting in $\tau_{xy}\sim161$ days. The total time duration of the light curves are 648 days.
The ICCF, computed again over a range from -50 to 100 days,
peaks around $\tau\sim0$ day, with $r_{\rm max}=0.97$ and $z_{\rm max}=2.15$.
Around $\tau\sim 0$ day, the cross-correlation coefficients exceed the 3$\sigma$ confidence region
($|z|\lesssim1.86$) of the null hypothesis. If we adopt $\sigma_z\sim0.62$ and $m=150/2\tau_{xy}\sim1$,
the probability distribution $p(z_{\rm max})$ given by Equation~(\ref{eqn_prmax2}) yields a probability of $\sim3\times10^{-4}$
for $z_{\rm max}>z_{\rm max, obs}$ in the null hypothesis.

We perform an additional test by combining a segment of the UVW2-band light curve of Fairall 9 (with a time shift)
and the H$\beta$ light curve of Mrk 335 to construct a new dataset. In the bottom panels of Figure~\ref{fig_tests},
we plot the mock light curves and the resulting ICCF over a time-lag range from -100 to 100 days.
We find that all the cross-correlation coefficients
lie well within the 2$\sigma$ confidence region. The ICCF peaks around $\tau\sim-50$ days with $r_{\rm max, obs}=0.66$ and
$z_{\rm max,obs}=0.79$. Regarding the probability distribution $p(z_{\rm max})$, the obtained $\tau_{xy}\sim39$
days yields a rough estimate of the parameter $m\sim200/2\tau_{xy}\sim2.6$. By adopting $\sigma_z\sim0.57$,
the probability for $z_{\rm max}>z_{\rm max, obs}$ is $\sim$0.18 in the null hypothesis, which is significantly
higher than that for Mrk~335 and Fairall~9. This is as expected since the mock data are indeed uncorrelated.
This test also illustrates that the peak cross-correlation coefficient is not an absolutely reliable criteria for cross-correlation
analysis.

In all three cases above, the ratio $\tau_{xy}/T_{\rm dur}$ ranges from about 0.13 to 0.26. Figure~\ref{fig_sigma} illustrates
that the variances are slightly overestimated by a factor of $\lesssim$1.5 using
the theoretical expressions in Section~\ref{sec_ccf}. Accordingly, the standard deviations
are overestimated by a factor of $\lesssim$1.2, indicating that the obtained
cross-correlation significances are not significantly affected.

\section{Conclusion and Discussion}\label{sec_conclusion}
In this work, we investigate the probability distributions of the cross-correlation coefficients and their peak
value for AGN light curves. Our analysis is grounded in a fundamental corollary
from stochastic time series theory, which dictates that for stationary light curves, the cross-correlation coefficient
of independent light curves asymptotically approaches a normal distribution,
with its variance expressible in terms of the auto-correlation functions of the individual light curves.
Using Monte Carlo simulations, we demonstrate that this property also holds for irregularly sampled AGN light curves.
We thereby develop a fast procedure to estimate the significance level of
the cross-correlation coefficient at a given time lag in the null hypothesis, namely,
the light curves are independent and uncorrelated.
We also derive an approximate expression for the probability distribution for
the peak cross-correlation coefficient of uncorrelated light curves.
As illustrative examples, we apply these results to reverberation mapping data for the broad-line region
of Mrk 335 and the accretion disk of Fairall 9 from the literature.
We incorporate all procedures developed in this work into a Python package \texttt{PyAT}, publicly available
at \url{https://github.com/LiyrAstroph/PyAT}.

Finally, it is worth stressing the following two remarks.

First, the theoretical estimate of $\sigma_z^2$ requires prior determination of the auto-correlation functions of the light curves.
In principle, this can be achieved via fitting a specific variability model, e.g.,
the widely used damped random walk model, which is described by two parameters,
the damping timescale and long-term standard deviation.
Previous studies showed that for the damped random walk model, the auto-correlation function
might not be reliably recovered when the time duration of the light curve is comparable to the typical damping timescale
(e.g., \citealt{Kozlowski2017, Hu2024}). In this case, the damping timescale is usually underestimated,
leading to a smaller $\sigma_z^2$ when adopting the inferred damping timescales.
This effect partially mitigates the overestimation of $\sigma^2_z$ derived from the true damping timescales, as
illustrated in Section~\ref{sec_bias}. Nevertheless, how to reliably infer the damping timescale
remains an open issue and warrants further exploration in future work.

Second, in AGN reverberation mapping analysis, we generally adopt a strong prior that
AGN light curves are correlated across different bands
from UV to optical, as well as between UV/optical continuum and broad emission lines. These correlations
have been extensively validated by observations (e.g., \citealt{Peterson2002, Edelson2019}).
Therefore, the above null-hypothesis test should be regarded as an approach for assessing the reliability of the
cross-correlation coefficients of the observed light curves (see also the discussion in \citealt{U2022}).
In addition, there is also not a strict threshold for the significance level to distinguish between correlated and
uncorrelated light curves.

\section*{Acknowledgements}
We thank the referee for the useful comments that improved the manuscript.
We acknowledge financial support from the National Key R\&D Program of China (2023YFA1607904 and 2021YFA1600404)
and from the National Natural Science Foundation of China (NSFC; 12521005).
Y.-R.L. acknowledges financial support from the NSFC through grant No. 12273041, the Youth Innovation Promotion Association CAS,
and the China-Chile Joint Research Fund (CCJRF2310).
J.-M.W. acknowledges financial support from the NSFC through grant No. 12333003.

\software{\texttt{PyAT} (\citealt{Li_PyAT})}

\appendix
In this appendix, we provide a derivation for Equation~(\ref{eqn_norm}). For convenience, we define the following denotations
\begin{equation}
 u_i = x_i-\bar x, ~~{\rm and}~~ v_i = y_i - \bar y,
\end{equation}
where
\begin{equation}
\bar x = \frac{1}{n}\sum_i^n x_i, ~~{\rm and}~~\bar y = \frac{1}{n}\sum_i^n y_i.
\end{equation}
From Equation~(\ref{eqn_r}), the cross-correlation coefficient is recast as
\begin{equation}
r = \frac{\sum_{i=1}^{n} u_iv_i}{\sqrt{\sum_{i=1}^nu_i^2\sum_{i=1}^{n}v_i^2}} = \frac{1}{n}\frac{\sum_{i=1}^{n} u_iv_i}{\sigma_x\sigma_y},
\end{equation}
where
\begin{equation}
\sigma_x = \left(\frac{1}{n}\sum_{i=1}^n u_i^2\right)^{1/2},~~{\rm and}~~\sigma_y = \left(\frac{1}{n}\sum_{i=1}^n v_i^2\right)^{1/2}.
\end{equation}
We note that the above expressions for $\sigma_x$ and $\sigma_y$ are biased (\citealt[Section 7.2]{Brockwell1991}),
however, the biases are negligible for large $n$, which is generally satisfied in AGN reverberation mapping analysis.
Because both $x$ and $y$ light curves are stationary, their variances $\sigma_x$ and $\sigma_x$ are time-independent.
It is easy to prove that the ensemble mean of $r$ is zero, i.e., $\langle r \rangle=0$,
where the brackets represent the ensemble average.
The ensemble variance of $r$ is
\begin{equation}
\sigma_r^2=\langle r r\rangle =  \frac{1}{n^2}\frac{\sum_{i, j} \left\langle u_iu_jv_iv_j\right\rangle}{\sigma_x^2\sigma_y^2},
\end{equation}
where both $i$ and $j$ run from 1 to $n$.
Assuming that $u_i$ and $v_j$ are uncorrelated,
\begin{eqnarray}
\sum_{i, j}\left\langle u_iu_jv_iv_j\right\rangle &=& \sum_i\left\langle u_i^2 v_i^2\right\rangle + \sum_{i\neq j}\left\langle u_iu_jv_iv_j\right\rangle\nonumber\\
&=& \sum_i \left\langle u_i^2 \right\rangle\left\langle v_i^2\right\rangle + \sum_{i\neq j} \left\langle u_iu_j\right\rangle \left\langle v_iv_j\right\rangle.
\end{eqnarray}
Again, because $x$ and $y$ light curves are stationary, we have
\begin{equation}
\left\langle u_i^2 \right\rangle = \gamma_x(0)=\sigma_x^2, ~~ \left\langle u_iu_j\right\rangle = \gamma_x(i-j),
\end{equation}
and
\begin{equation}
\left\langle v_i^2 \right\rangle = \gamma_y(0)=\sigma_y^2, ~~  \left\langle v_iv_j\right\rangle = \gamma_y(i-j),
\end{equation}
where $\gamma_x(k)$ and $\gamma_y(k)$ are the covariance functions of $x$ and $y$ light curves at a time lag of $k\Delta t$, respectively. With the above expressions, the ensemble variance of $r$ is given by
\begin{eqnarray}
\sigma_r^2 &=& \frac{1}{n^2}\left[ n + \sum_{i\neq j}\rho_x(i-j)\rho_y(i-j)\right]\nonumber\\
&=& \frac{1}{n}\left[1 + \sum_{1\leqslant|k|\leqslant n} \left(1-\frac{|k|}{n}\right)\rho_x(k) \rho_y(k)\right]\\
&=& \frac{1}{n}\left[ 1 + 2 \sum_{k=1}^{n}\left(1-\frac{k}{n}\right)\rho_x(k)\rho_y(k)\right],\nonumber
\end{eqnarray}
where
\begin{equation}
\rho_x(k) = \frac{\gamma_x(k)}{\sigma_x^2},~~\rho_y(k) = \frac{\gamma_y(k)}{\sigma_y^2},
\end{equation}
and the covariance function is an even function. According to Equation~(\ref{eqn_rz}),
the variances of $r$ and $z$ are approximately equal, namely, $\sigma_z^2\approx\sigma_r^2$.

\bibliography{refs.bib}{}
\bibliographystyle{aasjournalv7}
\end{document}